\documentclass[11pt]{article}
\usepackage{amssymb}
\usepackage{epsfig}

\newcommand{\BE}{\begin{equation}}
\newcommand{\EE}{\end{equation}}
\begin{document}
\begin{titlepage}

\vspace*{1mm}
\begin{center}

\LARGE
   {\LARGE{\bf The vacuum as a form of turbulent fluid:\\ motivations,
   experiments, implications }}

\vspace*{10mm} {\Large  M. Consoli$^a$, A. Pluchino$^{a,b}$, A. Rapisarda$^{a,b}$ and S.
Tudisco$^{c}$}
\vspace*{8mm}\\
{\large
a) Istituto Nazionale di Fisica Nucleare, Sezione di Catania \\
b) Dipartimento di Fisica e Astronomia dell' Universit\`a di Catania \\
c) Istituto Nazionale di Fisica Nucleare, Laboratori Nazionali del Sud, Catania\\}
\end{center}
\vspace*{5mm}
\begin{center}
{\bf Abstract}
\end{center}

\noindent Basic foundational aspects of both quantum theory and
relativity might induce to represent the physical vacuum as an
underlying highly turbulent fluid. By explicit numerical
simulations, we show that a form of statistically isotropic and
homogeneous vacuum turbulence is entirely consistent with the
present ether-drift experiments. In particular, after subtracting
known forms of disturbances, the observed stochastic signal requires
velocity fluctuations whose absolute scale is well described by the
average Earth's motion with respect to the Cosmic Microwave
Background. We emphasize that the existence of a genuine stochastic
ether drift could be crucial for the emergence of forms of
self-organization in matter and thus for the whole approach to
complexity.

\end{titlepage}

\section{Introduction}

According to the original Einstein view \cite{einstein}, the vacuum
could be regarded as trivially empty since Lorentz symmetry is an
exact symmetry of nature. In a Lorentzian approach
\cite{lorentz,poincare,electron}, on the other hand, there is an
underlying form of ether and Lorentz symmetry, rather than being
postulated from scratch, should be considered as an `emergent'
phenomenon. In spite of these deep conceptual differences, however,
it is far from obvious how to distinguish experimentally between
these two points of view. This type of conclusion was, for instance,
already clearly expressed by Ehrenfest in his lecture `On the crisis
of the light ether hypothesis' (Leyden, December 1912) as follows:
``So, we see that the ether-less theory of Einstein demands exactly
the same here as the ether theory of Lorentz. It is, in fact,
because of this circumstance, that according to Einstein's theory an
observer must observe exactly the same contractions, changes of
rate, etc. in the measuring rods, clocks, etc. moving with respect
to him as in the Lorentzian theory. And let it be said here right
away and in all generality. As a matter of principle, there is no
experimentum crucis between the two theories". This can be
understood since, independently of all interpretative aspects, the
basic quantitative ingredients, namely Lorentz transformations, are
the same in both formulations.

To understand this crucial aspect, one can use a very simple
argument. Suppose that the basic Lorentz transformations, rather
than originating from the relative motion of a pair of observers
$S'$ and $S''$, as in Einstein's relativity, might instead be
associated with their {\it individual} velocity parameters
$\beta'=v'/c$ and $\beta''=v''/c$  relatively to some preferred
frame $\Sigma$ \cite{bell,brown1,pla}. Still, due to the fundamental
group properties, the two frames $S'$ and $S''$ would also be
mutually connected by a Lorentz transformation with relative
velocity parameter \BE \beta_{\rm rel}=
{{\beta'-\beta''}\over{1-\beta'\beta''}}\equiv {{v_{\rm
rel}}\over{c}} \EE (we restrict for simplicity to one-dimensional
motion). This would produce a substantial quantitative equivalence
with Einstein's formulation for most standard experimental tests,
where one just compares the relative measurements of a pair of
observers. Hence, the importance of the ether-drift experiments
where one attempts to measure an absolute velocity.

At the same time, if the velocity of light $c_\gamma$ propagating in
the various interferometers coincides with the basic parameter $c$
entering Lorentz transformations, relativistic effects conspire to
make undetectable the individual $\beta'$, $\beta''$,...This means
that a null result of the ether-drift experiments should {\it not}
be automatically interpreted as a confirmation of Special
Relativity. As stressed by Ehrenfest, the motion with respect to
$\Sigma$ might remain unobservable, yet one could interpret
relativity ` \`a la Lorentz'. This could be crucial, for instance,
to reconcile faster-than-light signals with causality \cite{annals}
and thus provide a different view of the apparent non-local aspects
of the quantum theory \cite{scarani}.

However, to a closer look, is it really impossible to detect the
motion with respect to $\Sigma$? This possibility, which was
implicit in Lorentz' words \cite{electron} ``...it seems natural not
to assume at starting that it can never make any difference whether
a body moves through the ether or not..", may induce one to
re-consider the various issues and go deeper into the analysis of
the ether-drift experiments.

After this general premise, the scope of this paper is threefold.
First, in Sect.2, after comparing with basic foundational aspects of
both quantum physics and relativity, we will argue that the physical
vacuum could be represented as a random medium, similar to an
underlying turbulent fluid. Second, through Sects. 3$-$5, we will
show, by explicit numerical simulations, that a form of
statistically isotropic and homogeneous vacuum turbulence is
entirely consistent with the type of stochastic signal observed in
the present ether-drift experiments. In particular, after
subtracting known forms of disturbances, the observed signal is
consistent with velocity fluctuations whose absolute scale is fixed
by the average Earth's motion with respect to the Cosmic Microwave
Background. A definite confirmation (or refutation) of this result
should be obtained with the next generation of cryogenic
experiments. Finally, in Sect.6, in the conclusions, we will
emphasize that the detection of a genuine stochastic ether drift
could also be crucial to understand the emergence of forms of
self-organization in matter and thus for the whole approach to
complexity. In this sense, the ultimate implications of this
analysis could go far beyond the mere interpretation of relativity.

\section{The physical vacuum as a form of turbulent fluid}

In this section, we will list several different motivations that
might induce to represent the vacuum as a form of random medium
which resembles a turbulent fluid.

~~~~i) One could start by recalling that at the dawn of XX century
Lorentz symmetry was believed to emerge from an underlying ether
represented, by Thomson, Fitzgerald and others, as an incompressible
turbulent fluid ( a vortex `sponge') \cite{whittaker}. More
recently, the turbulent-ether model has been re-formulated by
Troshkin \cite{troshkin} (see also \cite{puthoff} and
\cite{tsankov}) in the framework of the Navier-Stokes equation and
by Saul  \cite{saul} by starting from Boltzmann's transport
equation. The main point of these hydrodynamic derivations is that,
due to the energy which is locally stored in the turbulent motion,
on a coarse-grained scale, a fluid can start to behave as an elastic
medium and thus support the propagation of transverse waves whose
speed $c_\gamma$ coincides with the average speed $c \equiv c_{\rm
turbulence}$ of the chaotic internal motion of the elementary fluid
constituents.

In this sense, the basic phenomenon of turbulence provides a
conceptual transition from fluid dynamics to a different realm of
physics, that of elasticity \footnote{The origin of this concept
could probably be searched into Hertz's mechanics \cite{gantmacher}
with his idea of microscopic, hidden motions whose kinetic energy is
actually the source of the forms of potential energy that we observe
in nature.}. This conclusion is also supported by the formal
correspondence \cite{marcinkowski,kosevicbook} (velocity potential
vs. displacement, velocity vs. distortion, vorticity vs. density  of
dislocations,...) that can be established between various systems of
dislocations in an elastic solid and vortex fields in a liquid. With
this transition the parameter $c$ acquires also the meaning of a
{\it limiting} speed for moving dislocations. This is due to the
behaviour of their elastic energy which increases proportionally to
$(1-v^2/c^2)^{-1/2}$. For this reason, dislocations have been
considered as a possible model for ordinary matter, see e.g.
refs.\cite{frank}$-$\cite{christov}.

This perspective is similar to starting from the basic equation that
determines the mutual variations of the energy $E$ and the linear
momentum ${\bf p}=M{\bf v}$ of a body \BE \label{classic}
{{dE}\over{dt}}= {\bf v}\cdot {{d(M{\bf v)}}\over{dt}} \EE and
allowing for a $v^2-$dependence in $M$ (see e.g. \cite{fey}). This
gives \BE \label{inter} dE= {{1}\over{2}}M dv^2 + v^2 dM \EE The
main point is that, if ordinary matter were interpreted in terms of
soliton-like excitations of an underlying turbulent ether, one now
disposes of the velocity parameter $c \equiv c_{\rm turbulence}$.
Then, by setting \footnote{As an example of this proportionality
relation, one can consider the case of quantum vortices (rotons)
within Landau's original quantum hydrodynamics. There, it is the
squared zero-point speed $c^2_{\rm zp}$ of the fluid constituents to
determine the proportionality relation between the energy gap
$E_{\rm roton}$, to produce vortical excitations, and their inertial
mass $M_{\rm roton}$ \cite{pla2012}.} $E\equiv c^2 M(v^2/c^2)$ , one
has ${{dE}\over{dv^2}}= c^2 {{dM}\over{dv^2}}$ and Eq.(\ref{inter})
becomes \BE {{dM}\over{dv^2}} \left( c^2 - v^2 \right)=
{{1}\over{2}}M \EE Therefore, for $dM/dv^2 > 0$, $c$ plays also the
role of a limiting speed and one finally obtains \BE \label{final}
E=Mc^2= {{M_0 c^2}\over{\sqrt{1-v^2/c^2} }}\EE On this basis, it
becomes natural to introduce linear transformations of the four
quantities $E/c$ and ${\bf p}=M{\bf v}$ that preserve the quadratic
combination $(E/c)^2 -{\bf p}^2=(M_0c)^2$ and thus, by starting from
a microscopic turbulent-ether scenario, Lorentz symmetry could also
be understood as an emergent phenomenon. In this interpretation, its
ultimate origin has to be searched in the very existence of $c$ and
thus in the deepest random fluctuations of the fluid velocity, with
time at each point and between different points at the same instant,
that characterize a state of fully developed turbulence and provide
a kinetic basis for the observed space-time symmetry \cite{pla2012}.

Notice that, once Lorentz symmetry is an emergent property, $c$ is
only a limiting speed for those soliton-like, collective modes that,
in an emergent interpretation, are taken as models of ordinary
matter, e.g. vortices, elastic dislocations...Thus there is nothing
wrong if the internal motion of the basic constituents takes place
at an average speed $c$. At the same time, on the coarse grained
scale which is accessible to physical rods and clocks, the basic
constituents appear, so to speak, `frozen' in the vacuum structure
and only their collective excitations are directly observable. This
means that, for the elementary ether constituents,
Eq.(\ref{classic}) is now solved by the standard non-relativistic
forms $E={{1}\over{2}}mv^2$ and ${\bf p}=m{\bf v}$, where $m$ is the
constituent constant mass.

~~~ii) This qualitative picture of the vacuum, as an underlying
random medium, also arises from alternative views of the quantum
phenomena as with stochastic electrodynamics
\cite{marshall}$-$\cite{cole} or Nelson's mechanics \cite{nelson}
(see \cite{chaos} for more details). The former is essentially the
classical Lorentz-Dirac theory \cite{dirac38} with new boundary
conditions where the standard vanishing field at infinity is
replaced by a vacuum, random radiation field. This field, considered
in a stationary state, is assumed to permeate all space and its
action on the particles impresses upon them a stochastic motion with
an intensity characterized by Planck's constant. In this way, one
can get insight into basic aspects of the quantum theory such as the
wave-like properties of matter, indeterminacy, quantization,... For
instance, in this picture, atomic stability would originate from
reaching that `quantum regime' \cite{puthoffH,cole} which
corresponds to a dynamic equilibrium between the radiation emitted
in the orbital motions and the energy absorbed in the highly
irregular motions impressed by the vacuum stochastic field. In this
sense, again, Lorentz' ether should not be thought as a stagnant
fluid (for an observer at rest) or as a fluid in laminar motion (for
an observer in uniform motion). Rather the ether should resemble a
fluid in a chaotic state, e.g. a fluid in a state of turbulent
motion. The same is true for Nelson's mechanics. Here, the idea of a
highly turbulent fluid emerges if one uses Onsager's original result
\cite{onsager} that in the zero-viscosity limit, i.e. infinite
Reynolds number, the fluid velocity field does not remain a
differentiable function \footnote{Onsager's argument relies on the
impossibility, in the zero-viscosity limit, to satisfy the
inequality $|{\bf v} ({\bf x} + {\bf l})- {\bf v}({\bf x})| < ({\rm
const.}) l^n$, with $n > 1/3$. Kolmogorov's theory \cite{kolmo}
corresponds to $n=1/3$. }. This provides a basis to expect that
``the Brownian motion in the ether will not be smooth
''\cite{nelson} and thus to consider the particular form of
kinematics which is at the basis of Nelson's stochastic derivation
of the Schr\"odinger equation.

~~~iii) At a more elaborate level, a qualitatively similar picture
is also obtained by representing relativistic particle propagation
from the superposition, at very short time scales, of
non-relativistic particle paths with different Newtonian mass
\cite{kleinert}. In this formulation, particles randomly propagate
(in the sense of Brownian motion) in a granular medium which thus
replaces the trivial empty vacuum \cite{jizba}. The essential
mathematical ingredient for this representation is the use of
`superstatistics' \cite{beck-cohen1,beck-cohen2}, intended as the
superposition of several statistical systems that operate at
different spatio-temporal scale, which is also known to provide a
very good description of fluid particle trajectories in high
Reynolds-number turbulence \cite{reynolds,beck}.

~~~iv) Finally, the idea of a fundamentally random vacuum is also
motivated by quantum-gravity. According to this view, space-time,
when resolved at very short distances, should exhibit quantum
fluctuations and thus appear to be `foamy' or `spongy' in the sense
of refs. \cite{wheeler1,hawking}.  This original idea has lead to a
very wide collection of ideas and intuitions including, for
instance, the holographic principle (see \cite{bousso} for a
review), possible deformations of Lorentz symmetry (Doubly Special
Relativity) \cite{amelino1,amelino2} or models of dark energy and
dark matter \cite{ng3}. At the same time, coupling light and matter
to a fluctuating metric leads to intrinsic limitations on the
measurement of lengths \cite{vandam,jaekel}, to violations of the
weak equivalence principle \cite{goklu} and to an effective
decoherence of quantum systems \cite{goklu1}. These effects can be
used to restrict the possible quantum gravity models by comparing
with the results of modern gravity-wave detectors
\cite{amelino3,amelino4} or with atomic interferometry \cite{goklu2}
or with the beat signal of two ultrastable optical resonators
\cite{braxmaier}. What is relevant here for our purpose is that, as
in the previous cases, the space-time foam of quantum gravity seems
also to resemble a turbulent fluid. This idea, originally due to
Wheeler \cite{wheeler1}, has been more recently exploited by Ng and
collaborators \cite{ng2,ng4} who have emphasized the close analogies
between holographic models of space-time foam and the limit of
turbulence for infinite Reynolds number. The main conclusion of
these rather formal derivations is that the metric fluctuations in
the holographic model, which give rise to length fluctuations
$\Delta l \sim l^{1/3} l_{\rm planck} ^{2/3}$, when compared with
those in moving fluids, can also be interpreted as a manifestation
of Kolmogorov's scaling law for velocity $\Delta v\sim l^{1/3}$
\cite{kolmo}.

Thus, summarizing, from the old ether view to the present
quantum-gravity models, there are several independent motivations to
represent the physical vacuum as an underlying turbulent fluid. One
could conclude that this non-trivial degree of convergence
originates from the fundamental nature of quantum gravity (e.g. from
the correspondence between the metric fluctuations in the
holographic model and Kolmogorov's scaling law). However, one could
also adopt the complementary point of view where instead the
ubiquitous phenomenon of turbulence plays from the very beginning
the most central role. In any case, it becomes natural to wonder
whether this type of vacuum medium could represent the preferred
reference frame of a Lorentzian approach and thus to look at the
results of the modern ether-drift experiments for experimental
checks. At the same time, the non-trivial interplay between
large-scale and small-scale properties of turbulent flows may induce
one to re-consider some assumptions adopted in the interpretation of
the data. These issues will be analyzed in detail in the following
three sections.

\section{The ether-drift experiments and the velocity of light}

As anticipated in the Introduction, the crucial issue in the context
of the ether-drift experiments concerns the value of $c_\gamma$, the
speed of light in the vacuum (measured on the Earth's surface). If
this coincides with the basic parameter $c$ entering Lorentz
transformations relativistic effects conspire to make undetectable
the individual $\beta'$, $\beta''$,... Therefore the only
possibility is that $c_\gamma$ and $c$ do not coincide {\it
exactly}, see e.g. \cite{uzan}. In this case, in fact, this mismatch
would show up through a tiny ether-drift effect $\delta \sim
\beta^2(c-c_\gamma)/c$.

This possibility was explored in ref.\cite{gerg} within the so
called emergent-gravity scenario \cite{barcelo1,barcelo2} where the
physical vacuum is modeled as a moving fluid with a small
compressibility. In this framework $(c-c_\gamma)/c$ was estimated
\cite{gerg} to be ${\cal O}(10^{-9})$, a value which is not ruled
out by the present experimental data. In fact the ether-drift, as
measured from the fractional beat signal between two vacuum optical
resonators \cite{applied,lammer}, gives $\delta \sim 10^{-15}$ and
thus could indicate a value $\beta^2 \sim 10^{-6}$, or an absolute
Earth's velocity of about 300 km/s, as for most cosmic motions. This
basic point can be easily checked by looking at Fig. 9(a) of
ref.\cite{crossed} where a typical sequence of 40 data collected at
regular steps of 1 second is reported (see our Fig.2 below)
\footnote{With respect to other articles, ref.\cite{crossed} has the
advantage to report the instantaneous raw data. The experiment also
adopts a sophisticated geometrical set-up where, to minimize all
possible asymmetries, the two optical cavities are obtained from the
same block of ULE (Ultra Low Expansion) material. As such, the
results of ref.\cite{crossed} will play an important role in our
analysis.}. As one can see, this instantaneous signal exhibits
random fluctuations of about $\pm 1$ Hz and this value, for the
given laser frequency $2.82\cdot 10^{14}$ Hz, might correspond to a
genuine ether-drift $\delta$ of about $\pm 3.5 \cdot 10^{-15}$. To
better appreciate this point, let us resume the various aspects
which are needed for the analysis of the experiments.

The basic concept in ether-drift experiments is the two-way velocity
of light in the vacuum $\bar{c}_\gamma (\theta)$. This is defined in
terms of the one-way velocity ${c}_\gamma (\theta)$ (which is not
unambiguously measurable) through the relation
\begin{eqnarray}
\label{rtwoway}
       \bar{c}_\gamma(\theta)&=&
       {{ 2  c_\gamma(\theta) c_\gamma(\pi + \theta) }\over{
       c_\gamma(\theta) + c_\gamma(\pi + \theta) }}
\end{eqnarray}
and could exhibit a non-zero anisotropy \BE \label{rrms}
    {{\Delta \bar{c}_\theta } \over{c}}=
    {{\bar{c}_\gamma(\pi/2 +\theta)- \bar{c}_\gamma (\theta)} \over
       {\langle \bar{c}_\gamma \rangle }} \neq 0 \EE
This theoretical concept is related to the measurable frequency
shift, i.e. the beat signal, $\Delta \nu$ of two optical resonators
\cite{applied,lammer} through the relation \BE \label{bbasic2}
     \delta(t)\equiv {{\Delta \bar{c}_\theta(t) } \over{c}} =
{{\Delta \nu^{\rm phys} (t)}\over{\nu_0}} \EE where $\nu_0$ is the
reference frequency of the two optical resonators and the suffix
``${\rm phys}$" indicates a hypothetical physical part of the
frequency shift after subtraction of all spurious effects.

As a possible theoretical framework for a non-zero anisotropy, we
shall concentrate on a scenario which introduces some difference
with respect to standard General Relativity and has a very simple
motivation: $\bar{c}_\gamma$ might differ from the basic parameter
$c$ entering Lorentz transformations due to gravitational effects.
To this end, as anticipated, one can consider the emergent-gravity
scenario \cite{barcelo1,barcelo2} where the space-time curvature
observed in a gravitational field becomes an effective phenomenon in
flat space, analogously to a hydrodynamic description of a moving
fluid on length scales which are much larger than the size of its
elementary constituents. In this perspective, gravity produces local
modifications of the basic space-time units which are known, see
e.g. \cite{feybook,dicke1}, to represent an alternative way to
introduce the concept of curvature \footnote{This point of view has
been vividly represented by Thorne in one of his books
\cite{thorne}: "Is space-time really curved ? Isn't it conceivable
that space-time is actually flat, but clocks and rulers with which
we measure it, and which we regard as perfect, are actually rubbery
? Might not even the most perfect of clocks slow down or speed up
and the most perfect of rulers shrink or expand, as we move them
from point to point and change their orientations ? Would not such
distortions of our clocks and rulers make a truly flat space-time
appear to be curved ? Yes". }. This scenario represents the simplest
modification of the standard picture which allows for a
non-vanishing anisotropy and gives the correct order of magnitude
$\delta \sim 10^{-15}$. As such, it will be adopted  in the rest of
this paper.

For the general problem of measuring the speed of light, one should
start, as in ref. \cite{gerg}, from the basic notion: the definition
of speed as (distance moved)/(time taken). To this end, one has to
choose some standards of distance and time and different choices can
give different answers. Therefore, we shall adopt the {\it same}
point of view of special relativity: the right space-time units are
those for which the two-way velocity of light in the vacuum
$\bar{c}_\gamma$, when measured in an inertial frame, coincides with
the basic parameter $c$ entering Lorentz transformations. However,
inertial frames are just an idealization. Therefore the appropriate
realization is to assume {\it local} standards of distance and time
such that the identification $\bar c_\gamma=c$ holds as an
asymptotic relation in the physical conditions which are as close as
possible to an inertial frame, i.e. {\it in a freely falling frame}
(at least by restricting to a space-time region small enough that
tidal effects of the external gravitational potential $U_{\rm
ext}(x)$ can be ignored). This is essential to obtain an operative
definition of the otherwise unknown parameter $c$. With these
premises, light propagation for an observer $S'$ sitting on the
Earth's surface can be described with increasing degrees of
approximations \cite{gerg,chaos}:

~~~i) In a zeroth-order approximation, $S'$ is considered a freely
falling frame. This amounts to assume $c_\gamma=c$ so that, given
two events which, in terms of the local space-time units of $S'$,
differ by $(dx, dy, dz, dt)$, light propagation is described by the
condition (ff='free-fall') \BE\label{zero1} (ds^2)_{\rm ff}=c^2dt^2-
(dx^2+dy^2+dz^2)=0~\EE ~~~ii) However, is really the Earth a
freely-falling frame ? To a closer look, in fact, an observer  $S'$
placed on the Earth's surface can only be considered as a
freely-falling observer up to the presence of the Earth's
gravitational field. Its inclusion leads to tiny deviations from the
standard Eq.(\ref{zero1}). These can be estimated by considering
$S'$ as a freely-falling observer (in the same external
gravitational field described by $U_{\rm ext}(x)$) that however is
also carrying on board a heavy object of mass $M$ (the Earth's mass
itself) that affects the effective local space-time structure, see
Fig.1 of ref.\cite{chaos}. To derive the required correction, let us
again denote by ($dx$, $dy$, $dz$, $dt$) the local space-time units
of the freely-falling observer $S'$ in the limit $M=0$ and by
$\delta U$ the extra Newtonian potential produced by the heavy mass
$M$ at the experimental set up where one wants to describe light
propagation. In a flat-space interpretation, light propagation for
the $S'$ observer can then be described by the condition
 \BE\label{iso}(ds^2)_{\rm \delta U} ={{c^2d\hat{t}
^2}\over{{\cal N}^2 }}- (d\hat{x}^2+d\hat{y}^2+d\hat{z}^2)=0~\EE
where, to first order in $\delta U$, the space-time units
($d\hat{x}$, $d\hat{y}$, $d\hat{z}$, $d\hat{t}$) are related to the
corresponding ones ($dx$, $dy$, $dz$, $dt$) for $\delta U=0$ through
an overall re-scaling factor \BE \label{lambda} \lambda= 1+{{|\delta
U|}\over{c^2}} \EE and we have also introduced a vacuum refractive
index \footnote{A general isotropic metric $(A,-B,-B,-B)$ depends on
two functions which, in a flat-space picture, can be interpreted in
terms of an overall re-scaling of the space-time units and of a
refractive index. Since physical units of time scale as inverse
frequencies, and the measured frequencies $\hat \omega$ for $\delta
U \neq 0$ are red-shifted when compared to the corresponding value
$\omega$ for $\delta U = 0$, this fixes the value of $\lambda$.
Furthermore, independently of the specific underlying mechanisms,
the two functions $A$ and $B$ can be related through the general
requirement $AB=1$ which expresses the basic property of light of
being, at the same time, a corpuscular and undulatory phenomenon
\cite{ultraweak}. This fixes the value of ${\cal N}$.
}\BE\label{refractive1}{\cal N}= 1+2{{|\delta U|}\over{c^2}} \EE
Therefore, to this order, light is formally described as in General
Relativity where one finds the weak-field, isotropic form of the
metric \BE\label{gr} (ds^2)_{\rm GR}=c^2dT^2(1-2{{|U_{\rm
N}|}\over{c^2}})- (dX^2+dY^2+dZ^2)(1+2{{|U_{\rm
N}|}\over{c^2}})\equiv c^2 d\tau^2 - dl^2\EE In Eq.(\ref{gr}) $U_N$
denotes the Newtonian potential and ($dT$, $dX$, $dY$, $dZ$)
arbitrary coordinates defined for $U_{\rm N}=0$. Finally, $d\tau$
and $dl$ denote the elements of proper time and proper length in
terms of which, in General Relativity, one would again deduce from
$ds^2=0$ the same universal value $c={{dl}\over{d\tau}}$. This is
the basic difference with Eqs.(\ref{iso})-(\ref{refractive1}) where
the physical unit of length is $\sqrt
{d\hat{x}^2+d\hat{y}^2+d\hat{z}^2}$, the physical unit of time is
$d\hat{t}$ and  instead a non-trivial refractive index ${\cal N}$ is
introduced. For an observer placed on the Earth's surface, its value
is \BE \label{refractive}{\cal N}- 1 \sim {{2G_N M}\over{c^2R}} \sim
1.4\cdot 10^{-9}\EE where $G_N$ is Newton's constant and  $M$ and
$R$ are respectively the Earth's mass and radius.

~~~iii) Differently from General Relativity, in a flat-space
interpretation with re-scaled units ($d\hat{x}$, $d\hat{y}$,
$d\hat{z}$, $d\hat{t}$) and ${\cal N}\neq 1$, the speed of light in
the vacuum $c_\gamma$ no longer coincides with the parameter $c$
entering Lorentz transformations. Therefore, as a general
consequence of Lorentz transformations, an isotropic propagation as
in Eq.(\ref{iso}) can only be valid for a special state of motion of
the Earth's laboratory. This provides the operative definition of a
preferred reference frame $\Sigma$ while for a non-zero relative
velocity ${\bf V}$  one expects off diagonal elements $g_{0i}\neq 0$
in the effective metric and a tiny light anisotropy. As shown in
Ref.\cite{gerg}, to first order in both $({\cal N}- 1)$ and $V/c$
one finds \BE g_{0i}\sim 2({\cal N}- 1){{V_i}\over{c}} \EE These off
diagonal elements can be imagined as being due to a directional
polarization of the vacuum induced by the now moving Earth's
gravitational field and express the general property \cite{volkov}
that any metric, locally, can always be brought into diagonal form
by suitable rotations and boosts. In this way, by introducing
$\beta=V/c$, $\kappa=( {\cal N}^2 -1)$ and the angle $\theta$
between ${\bf V}$ and the direction of light propagation, one finds,
to ${\cal O}(\kappa)$ and ${\cal O}(\beta^2)$, the one-way velocity
\cite{gerg}
\BE \label{oneway}
       c_\gamma(\theta)= {{c} \over{{\cal N}}}~\left[
       1- \kappa \beta \cos\theta -
       {{\kappa}\over{2}} \beta^2(1+\cos^2\theta)\right]
\EE and a two-way velocity of light
\begin{eqnarray}
\label{twoway}
       \bar{c}_\gamma(\theta)&=&
       {{ 2  c_\gamma(\theta) c_\gamma(\pi + \theta) }\over{
       c_\gamma(\theta) + c_\gamma(\pi + \theta) }} \nonumber \\
       &\sim& {{c} \over{{\cal N}}}~\left[1-\beta^2\left(\kappa -
       {{\kappa}\over{2}} \sin^2\theta\right) \right]
\end{eqnarray}
This allows to define the RMS \cite{robertson,mansouri} anisotropy
parameter ${\cal B }$ through the relation \BE \label{rms}
    {{\Delta \bar{c}_\theta } \over{c}}=
    {{\bar{c}_\gamma(\pi/2 +\theta)- \bar{c}_\gamma (\theta)} \over
       {\langle \bar{c}_\gamma \rangle }} \sim{\cal B }
       {{V^2 }\over{c^2}} \cos(2\theta) \EE
with
\BE \label{rmsb}
       |{\cal B }|\sim {{\kappa}\over{2}}\sim {\cal N}-1
\EE
From the previous analysis, by replacing the value of the refractive
index Eq.(\ref{refractive}) and adopting, as a rough order of
magnitude, the typical value of most cosmic motions $V\sim$ 300
km/s, one expects a tiny fractional anisotropy \BE \label{averani}
        {{\langle\Delta \bar{c}_\theta \rangle} \over{c}} \sim
       |{\cal B }|{{V^2 }\over{c^2}} ={\cal O}(10^{-15}) \EE
that could finally be detected in the present, precise ether-drift
experiments.

\section{The experiments in more details}

Let us now consider in more detail the experimental aspects. To
increase the statistics, the present experiments exhibit rotating
optical resonators. In this case, the relative frequency shift for a
symmetric set up can be expressed as \BE \label{basic2}{{\Delta
\nu^{\rm phys} (t)}\over{\nu_0}} = 2{S}(t)\sin 2\omega_{\rm rot}t +
      2{C}(t)\cos 2\omega_{\rm rot}t
\EE where $\omega_{\rm rot}$ is the rotation frequency of the
apparatus. The overall factor of two on the right hand side of the
above equation is needed to correctly normalize the measured shifts
in terms of the functions $S(t)$ and $C(t)$ extracted from the
non-symmetric apparatus of ref.\cite{peters} \footnote{In the
non-symmetric apparatus of ref.\cite{peters} one measures the
combination $\bar{c}_\gamma(0)- \bar{c}_\gamma (\theta)$. On the
other hand, in a fully symmetric apparatus one measures the other
combination $ \bar{c}_\gamma(\pi/2 +\theta)- \bar{c}_\gamma
(\theta)$. Apart from a constant offset, by using Eq.(\ref{twoway}),
the angular dependence of the two expressions differs by a relative
factor of two.}. Notice also that in some articles the function
$S(t)$ is denoted as $B(t)$.

In this framework, the existence of possible time modulations of the
signal that might be synchronous with the Earth's rotation has
always represented a crucial ingredient for the analysis of the
data. This expectation derives from a model where one assumes a {\it
fixed} preferred frame $\Sigma$. Then, for short-time observations
of 1-2 days, the time dependence of a hypothetical physical signal
can only be due to (the variations of the projection of the Earth's
velocity ${\bf V}$ in the interferometer's plane caused by) the
Earth's rotation. In this case, the two functions $S(t)$ and $C(t)$
admit the simplest Fourier expansion \cite{peters} ($t'=\omega_{\rm
sid}t$ is the sidereal time of the observation in degrees)  \BE
\label{amorse1}
      {S}(t) = S_0 +
      {S}_{s1}\sin t' +{S}_{c1} \cos t'
       + {S}_{s2}\sin(2t') +{S}_{c2} \cos(2t')
\EE \BE \label{amorse2}
      {C}(t) = {C}_0 +
      {C}_{s1}\sin t' +{C}_{c1} \cos t'
       + {C}_{s2}\sin(2 t') +{C}_{c2} \cos(2 t')
\EE
with {\it time-independent} $C_k$ and $S_k$ Fourier coefficients.

This theoretical framework, accepted so far by all experimental
groups, leads to average the various $C_k$ and $S_k$ obtained from
fits performed during a 1-2 day observation period. By further
averaging over many short-period experimental sessions, the data
support the general conclusion \cite{joint,newberlin,schillernew}
that, although the typical instantaneous $S(t)$ and $C(t)$ are
indeed ${\cal O}(10^{-15})$, the global averages $(C_k)^{\rm avg}$
and $(S_k)^{\rm avg}$ for the Fourier coefficients are much smaller,
at the level ${\cal O}(10^{-17})$, and, with them, the derived
parameters entering the phenomenological SME \cite{sme,sme3} and RMS
 models.

However, there might be different types of ether-drift where the
straightforward parameterizations Eqs.(\ref{amorse1}),
(\ref{amorse2}) and the associated averaging procedures are {\it
not} allowed. In fact, before assuming any definite theoretical
scenario, one should first ask: if light were really propagating in
a physical medium, an ether, and not in a trivial empty vacuum, how
should the motion of (or in) this medium be described? Namely, could
this relative motion exhibit variations that are {\it not} only due
to known effects as the Earth's rotation and orbital revolution? The
point is that, by representing the physical vacuum as a fluid, the
standard assumption of smooth sinusoidal variations of the signal,
associated with the Earth's rotation (and its orbital revolution),
corresponds to assume the conditions of a pure laminar flow
associated with simple regular motions. Instead, by adopting the
model of the vacuum as an underlying turbulent fluid, there might be
other forms of time modulations. In this alternative scenario, the
same basic experimental data might admit a different interpretation
and a definite instantaneous signal $\Delta \nu (t)\neq 0$ could
become consistent with $(C_k)^{\rm avg} \sim (S_k)^{\rm avg}\sim 0$.

To discuss this alternative scenario, it is convenient to first
re-write Eq.(\ref{basic2}) as \BE \label{basic3} {{\Delta \nu^{\rm
phys} (t)}\over{\nu_0}} = 2A(t)\cos (2\omega_{\rm rot}t
-2\theta_0(t)) \EE where \BE \label{interms}
C(t)=A(t)\cos2\theta_0(t)~~~~~~~~S(t)=A(t)\sin2\theta_0(t)\EE so
that \BE A(t)= \sqrt{S^2(t) +C^2(t)} \EE Here $\theta_0(t)$
represents the instantaneous direction of a hypothetical ether-drift
effect in the x-y plane of the interferometer (counted by convention
from North through East so that North is $\theta_0=0$ and East is
$\theta_0=\pi/2$). By also introducing the magnitude $v=v(t)$ of the
{\it projection} of the full ${\bf V}$, such that \BE
v_x(t)=v(t)\cos\theta_0(t)~~~~~~~~~~~~~~~v_y(t)=v(t)\sin\theta_0(t)
\EE we obtain the
  theoretical relations \cite{gerg}
\BE \label{amplitude1}
       A(t)= {{1}\over{2}}|{\cal B }| {{v^2(t) }\over{c^2}}
\EE and \BE \label{amplitude10}
       C(t)= {{1}\over{2}}{\cal B }~ {{v^2_x(t)- v^2_y(t)  }
       \over{c^2}}~~~~~~~~~~~~~~
       S(t)= {{1}\over{2}}{\cal B } ~{{2v_x(t)v_y(t)  }\over{c^2}}
\EE where ${\cal B }$ is the anisotropy parameter Eq.(\ref{rms}). In
the forthcoming section we shall produce a numerical simulation by
assuming a model of turbulent flow for the velocity components
$v_x(t)$ and $v_y(t)$ and computing $|{\cal B }|$ through
Eqs.(\ref{rmsb}) and (\ref{refractive}).

\section{Numerical simulation of a physical, stochastic component}

Before trying to simulate a physical stochastic component of the
signal, to obtain the correct normalization, we should first
subtract from the existing data the known spurious effects. To
obtain a precise statistical indicator we shall consider the root
square of the Allan variance (RAV) for an integration time $\tau
\sim 1$ second which we'll take as our definition of instantaneous
signal. In fact, for the considered laser frequency $\nu_0 \sim
2.82\cdot 10^{14}$ Hz, our model predicts typical frequency shifts
$\Delta \nu \lesssim $ 1 Hz so that, when looking for a beat signal,
it only makes sense to compare with sequences of data collected at
time steps of 1 second or larger.

The RAV describes the time dependence of an arbitrary function
$z=z(t)$ which can be sampled over time intervals of length $\tau$.
In this case, by defining \BE {\overline z}(t_i;\tau)={{1}\over{\tau
}}\int^{t_i+\tau }_{t_i}dt~z(t)\equiv {\overline z}_i \EE one
generates a $\tau-$dependent distribution of ${\overline z}_i$
values. In a large time interval $\Lambda= M\tau$, the RAV is then
defined as \BE{\rm RAV}(\tau)= \sqrt{\sigma^2(z,\tau) } \EE where
\BE \sigma^2(z,\tau)= {{1}\over{2M }}\sum^{M}_{i=1} \left({\overline
z}_i-{\overline z}_{i+1} \right)^2 \EE Now, for the non-rotating set
up, the RAV of the frequency shift for $\tau \sim 1$ second was
determined \cite{crossed} to be 0.8 Hz ($2.8\cdot 10^{-15}$ in
dimensionless units) \footnote{We tried to obtain an analogous
indication from the other experiment of ref.\cite{newberlin}.
However, for $\tau \sim1$ second, it is not so easy to determine the
value of the RAV. In fact, by inspection of their figure 2, in the
narrow range from $\tau= 0.8$ seconds to $\tau= 1$ second, the data
for the non-rotating set-up (the red dots) exhibit a very steep,
sizeable decrease from about $2.8\cdot 10^{-15}$ down to $1.4\cdot
10^{-15}$.} and found much larger than the corresponding
disturbances in the individual resonators (typically about 0.02-0.03
Hz). The only exception is the possible effect of thermal
disturbances in the mirrors and the spacers of the optical
resonators. This particular component should be independent of the
integration time and, for ULE optical resonators, on the basis of
the results of ref.\cite{numata}, was estimated in
ref.\cite{inglesi} to be about $1.15\cdot 10^{-15}$ in dimensionless
units. Therefore, for a laser frequency $\nu_0=2.82\cdot 10^{14}$
Hz, we would expect RAV(thermal-noise)$\sim 0.32$ Hz. It is
questionable how to subtract this effect  from the full measured
value 0.8 Hz. One might argue that, if the physical signal has also
a stochastic nature, one should subtract quadratically. This would
give \BE {\rm RAV(physical, \tau \sim 1~
second)=\sqrt{(0.8)^2-(0.32)^2}\sim 0.73 ~Hz}\EE Instead, we shall
adopt the more conservative attitude of subtracting linearly, i.e.
\BE {\rm RAV(physical, \tau\sim  1~ second)= 0.8 ~Hz -0.32 ~Hz= 0.48
~Hz}\EE or $1.7\cdot 10^{-15}$ in dimensionless units. Since for a
symmetric non-rotating set-up the physical signal is simply
$2C(t)\nu_0$, we conclude that there is a potentially important
contribution to $C(t)$ which corresponds to a stochastic signal with
an Allan variance of about $8.5\cdot10^{-16}$ for $\tau \sim 1$
second. This value will be the basic input for our simulation.

Let us now return to Eqs.(\ref{amplitude10}) and assume for the
velocity components $v_x(t)$ and $v_y(t)$ a model of turbulent flow.
This could be done in many different ways.  Here we shall restrict
to the simplest case of a turbulence which, in a wide range of
scales, appear statistically isotropic and homogeneous
\footnote{This picture reflects the basic Kolmogorov theory
\cite{kolmo} of a fluid with vanishingly small viscosity.}. To
describe the temporal pattern of the signal, we shall follow
ref.\cite{fung} where velocity flows, in statistically isotropic and
homogeneous 3-dimensional turbulence, are generated by unsteady
random Fourier series. The perspective is that of an observer moving
in the turbulent fluid who wants to simulate the two components of
the velocity in his x-y plane at a given fixed location in his
laboratory. This leads to the general expressions \BE \label{vx}
v_x(t)= \sum^{\infty}_{n=1}\left[
       x_n(1)\cos \omega_n t + x_n(2)\sin \omega_n t \right] \EE
\BE \label{vy} v_y(t)= \sum^{\infty}_{n=1}\left[
       y_n(1)\cos \omega_n t + y_n(2)\sin \omega_n t \right] \EE
where $\omega_n=2n\pi/T$, T being a time scale which represents a
common period of all stochastic components. In our simulation we
have fixed the typical value $T=T_{\rm day}$= 24 hours. However, we
have also checked with a few runs that the statistical distributions
of the various quantities do not change substantially if we vary $T$
in the rather wide range $0.1~T_{\rm day}\leq T \leq 10~T_{\rm
day}$.

The coefficients $x_n(i=1,2)$ and $y_n(i=1,2)$ are random variables
with zero mean. They have the physical dimension of a velocity and
we shall denote by $[-\tilde v,\tilde v]$ the relevant interval of
these parameters. In terms of $\tilde v$ the quadratic mean values
can be expressed as \BE \langle x^2_n(i=1,2)\rangle=\langle
y^2_n(i=1,2)\rangle={{{\tilde v}^2 }\over{3 ~n^{2\eta}}} \EE  for
the uniform probability model (within the interval $[-\tilde
v,\tilde v]$) which we have chosen for our simulations. Finally, the
exponent $\eta$ controls the power spectrum of the fluctuating
components. For our simulation, between the two values $\eta=5/6$
and $\eta=1$ reported in ref.\cite{fung}, we have chosen $\eta=1$
which corresponds to the point of view of an observer moving in the
fluid.

Thus, within this simple model for the stochastic signal, $\tilde v$
is our only free parameter and will be fixed by imposing that the
generated $C-$values give a RAV of $8.5\cdot10^{-16}$ for
integration time $\tau= 1$ second. By taking into account the
typical variation of the results, due to both the truncation of the
Fourier modes and the dependence on the random sequence, this
constraint gives a range $\tilde v\sim (332\pm 10)$ km/s which,
remarkably, has a definite counter part in the known Earth's motion
with respect to the Cosmic Microwave Background (CMB). In fact, it
coincides exactly with the daily average of the projection $\sqrt
{\langle v^2 \rangle}\sim 332$ km/s in the interferometer's plane
for an apparatus at the latitude of the laboratories in
Berlin-D\"usseldorf. This can be checked by using the relation
\cite{gerg} \BE \langle v^2 \rangle= V^2
       \left(1- \sin^2\gamma\cos^2\chi
       - {{1}\over{2}} \cos^2\gamma\sin^2\chi \right) \EE
and setting $V=$ 370 km/s, angular declination $\gamma\sim -6$
degrees and co-latitude $\chi\sim 38$ degrees.

\begin{figure}
\begin{center}
\epsfig{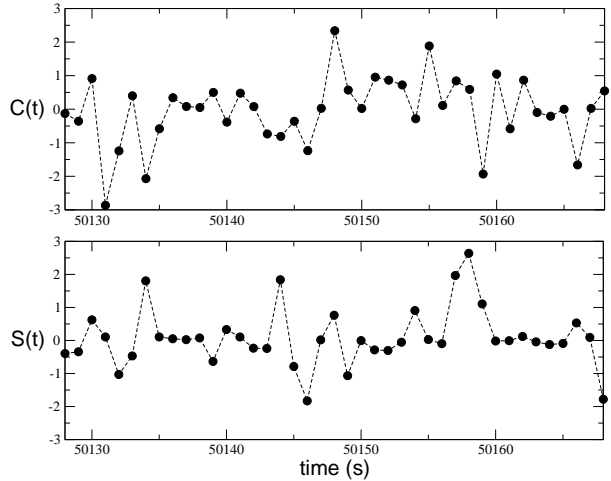}
\end{center}
\caption{ {\it A simulation of the instantaneous values of the $C$
and $S$ functions, in units $10^{-15}$, as obtained from a typical
sequence of 40 seconds. The combination $2C(t) \nu_0$ gives the
frequency shift for a symmetric non-rotating set up. Its general
trend should be compared with Fig.2 below from ref.\cite{crossed}.
In our simulation the effect of thermal noise has been preliminarily
subtracted out }. } \label{Fig.1}
\end{figure}

After these preliminaries, the results of our numerical simulation
can be illustrated by starting from the building blocks of our
scheme, namely the instantaneous values ${C}_i=C(t_i)$ and
${S}_i=S(t_i)$ of the $C$ and $S-$functions that determine the
frequency shift Eq.(\ref{basic2}). In Fig.1 we report a typical
sequence of 40 values of these functions. In particular the
combination $2C(t) \nu_0$ gives the frequency shift for a symmetric
non-rotating set up. The resulting general trend should be compared
with the experimental signal from ref.(\cite{crossed}) reported in
our Fig.2. The experimental frequency shifts are somewhat larger due
to the effect of thermal noise which has been preliminarily
subtracted out in our simulation.

\begin{figure}
\begin{center}
\epsfig{figure=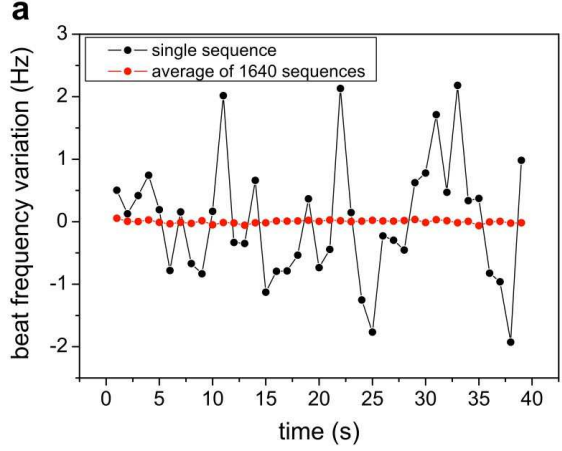,width=8truecm,angle=0}
\end{center}
\caption{ {\it The experimental signal, for a symmetric non-rotating
set up, reported in Fig.9(a) of ref.\cite{crossed} (courtesy Optics
Communications). For the given laser frequency $\nu_0=2.82\cdot
10^{14}$ Hz a frequency shift $\Delta \nu=\pm 1$ Hz corresponds to a
dimensionless ratio $\Delta \nu/\nu_0$ of about $\pm 3.5 \cdot
10^{-15}$ }. } \label{Fig.2}
\end{figure}

In terms of these basic quantities, one can construct a first type
of averages over a time scale $\tau\equiv N$ seconds \BE {\overline
C}(t_i;N)=
{{1}\over{N}}\sum^{i+N-1}_{n=i}{C}_n~~~~~~~~~~~~~~~~~~{\overline
S}(t_i;N)= {{1}\over{N}}\sum^{i+N-1}_{n=i}{S}_n \EE so that
${\overline C}(t_i;1)=C_i$ and ${\overline S}(t_i;1)=S_i$. This
first type of averaging is essential to compare with experiments
where the $C$ and $S-$functions are always determined after
averaging the basic instantaneous data over times $\tau$ in the
typical range 40-400 seconds. With these auxiliary quantities,
collected during a large time scale $\Lambda=M\tau$, one can form a
statistical distribution and determine mean values \BE \langle
C\rangle_\tau={{1}\over{M}}\sum^{M}_{i=1}{\overline
C}(t_i;\tau)~~~~~~~~~~~~~~~~~~\langle
S\rangle_\tau={{1}\over{M}}\sum^{M}_{i=1}{\overline S}(t_i;\tau) \EE
and variances \BE \sigma^2_C(\tau)=\sum^{M}_{i=1}{{ \left({\overline
C}(t_i;\tau)-\langle C\rangle_\tau\right)^2   }\over{M -1}}
~~~~~~~~~~~~\sigma^2_S(\tau)=\sum^{M}_{i=1}{{ \left({\overline
S}(t_i;\tau)-\langle S\rangle_\tau\right)^2   }\over{M -1}} \EE  We
report in Fig.3, for $\tau=1$ second, the distribution functions of
the simulated $\overline{C}$  and $\overline{S}$ values (panels (a)
and (b)).
\begin{figure}
\begin{center}
\epsfig{figure=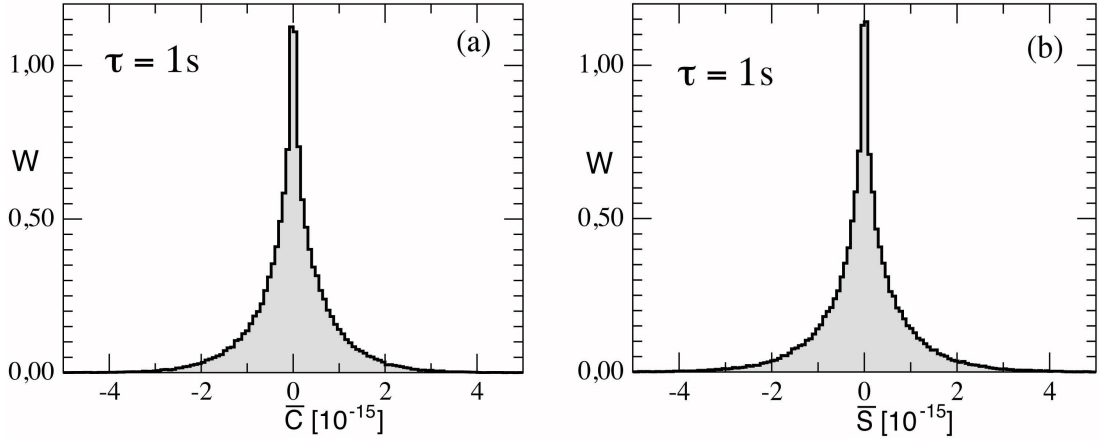,width=15truecm,angle=0}
\end{center}
\caption{ {\it We show, see (a) and (b), the histograms $W$ of the
simulated ${\overline C}$ and ${\overline S}$ values, in units
$10^{-15}$, for $\tau=1$ second. The vertical normalization is to a
unit area. The mean values are $\langle C\rangle_\tau =-1.1 \cdot
10^{-18}$, $\langle S\rangle_\tau =-1.9 \cdot 10^{-18}$ and the
standard deviations $\sigma_C(\tau)=8.5 \cdot 10^{-16}$,
$\sigma_S(\tau)=9.4\cdot 10^{-16}$. The total statistics correspond
to a time $\Lambda=M\tau=$86400 seconds. }}
\label{Fig.3}
\end{figure}
Notice that these distributions are clearly very different
from a Gaussian shape. This kind of behavior is known to
characterize probability distributions in turbulent flow at small
time scales (see e.g. \cite{sreenivasan,beck}).

By starting to average the instantaneous values, the statistical
distributions of the simulated ${\overline C}$ and ${\overline S}$ tend to
assume a gaussian shape. This is already evident from about
$\tau=5-6$ seconds. In Fig. 4 we show the two distributions for
$\tau=40$ seconds.

\begin{figure}
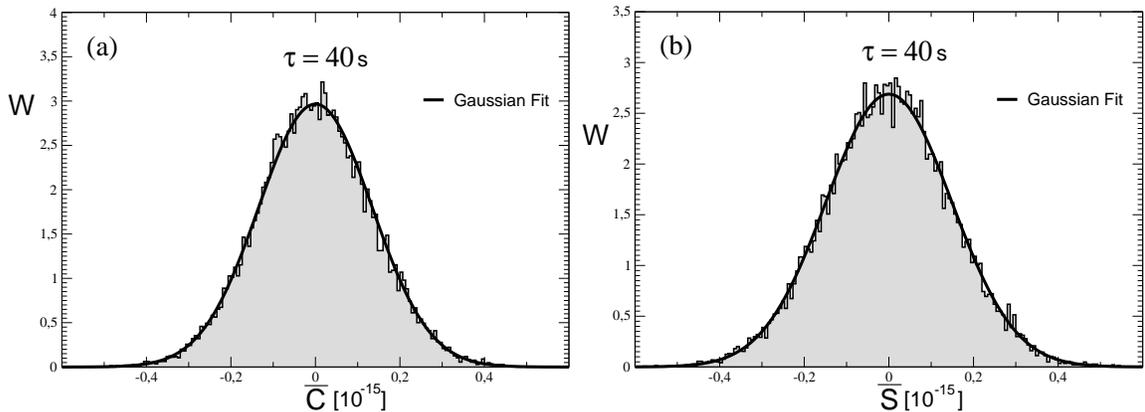

\begin{center}
\epsfig{figure=FFig3a.eps,width=7.5truecm,angle=0}
\epsfig{figure=FFig3b.eps,width=7.5truecm,angle=0}
\end{center}
\caption{ {\it The histograms $W$ of the simulated ${\overline C}$
and ${\overline S}$ values, in units $10^{-15}$, and the
corresponding gaussian fits for $\tau=40$ seconds. The vertical
normalization is to a unit area. The mean values are $\langle
C\rangle_\tau =-9 \cdot 10^{-19}$, $\langle S\rangle_\tau =-5 \cdot
10^{-19}$ and the standard deviations $\sigma_C(\tau)=1.34 \cdot
10^{-16}$, $\sigma_S(\tau)=1.48\cdot 10^{-16}$. The total statistics
correspond to a time $\Lambda=M\tau=$864000 seconds}. }
\label{Fig.4}
\end{figure}

As it might be expected, for all $\tau$ the statistical averages $
\langle C\rangle_\tau$ and $ \langle S\rangle_\tau$ are vanishingly
small in units of the typical instantaneous signal ${\cal
O}(10^{-15})$ and any non-zero value has to be considered as
statistical fluctuation. The standard deviations, on the other hand,
have definite values and exhibit a clear $1/\sqrt{\tau}$ trend so
that, to good approximation, one can express \BE \sigma_C(\tau)\sim
{{8.5\cdot 10^{-16}
 }\over{\sqrt{\tau ({\rm sec}) }}}~~~~~~~~~~~~~~~~~~ \sigma_S(\tau)\sim {{9.4\cdot
10^{-16}
 }\over{\sqrt{\tau ({\rm sec}) }}}\EE By keeping
 $\tilde v$ fixed at 332 km/s, the above two values for
$\tau=1$ second have an uncertainty of about $5\%$ which reflects
the typical variation of the results due to both the truncation of
the Fourier modes and the dependence on the random sequence.

Notice that our model predicts a monotonic decrease of the
dispersion of the data by increasing the averaging time $\tau$ and,
therefore, does not reproduce the linear increase of the Allan
variance which is seen, in all present room temperature experiments,
above about $\tau=100$ seconds. This is usually believed to be a
spurious thermal effect which, by the way, was also found in the
classical ether-drift experiments \footnote{To this end, one can
look at the original Michelson-Morley data, Am. J. Sci. 34 (1887)
333. As explained in Miller's  review article (see D. C. Miller,
Rev. Mod. Phys. 5 (1933) 203) the fringe shifts were obtained after
first correcting the data for the observed linear thermal drift.
This was producing a difference between the first reading and the
final reading obtained after a complete rotation of the
interferometer. If this correction were not implemented, no
meaningful interpretation of the classical ether-drift experiments
can be obtained.}. For this reason, the present limits on Lorentz
invariance refer crucially to the short-term stability of the
resonators. This thermal interpretation is also in agreement with
the cryogenic experiment of ref.\cite{mueller} where the Allan
variance (in the quiet phase between two refillings of the tank of
liquid helium) was found to exhibit a monotonic decrease up to about
$\tau=250$ seconds. It remains to be seen how far this decreasing
trend will be extended by the forthcoming generation of experiments
with cryogenic sapphire resonators \cite{upgrading} that are
expected to have a short-time stability of a few $10^{-18}$. Thus it
will be possible to obtain a precise check of our predictions. In
particular, the typical instantaneous signal should be about 100
times larger than the experimental sensitivity and the distributions
of the ${\overline C}(t_i;\tau)$ and ${\overline S}(t_i;\tau)$, for
$\tau=$ 100 seconds, should extend up to values which are still 10
times larger.

\section{Summary and outlook}

The ether-drift experiments play a fundamental role for our
understanding of relativity. In fact, so far, they are the only
known experiments which, in principle, can distinguish Einstein's
interpretation from the Lorentzian point of view with a preferred
reference frame $\Sigma$. Up to now, the interpretation of the data
has been based on a theoretical model where all type of signals that
are not synchronous with the Earth's rotation tend to be considered
as spurious instrumental noise and no particular effort is made to
understand if there could be genuine physical effects which do not
fit within the adopted scheme.

However, there is a logical gap which has been missed so far. Even
though the relevant Earth's cosmic motion corresponds to that
indicated by the anisotropy of the CMB ($V\sim $370 km/s, angular
declination $\gamma\sim -6$ degrees, and right ascension $\alpha
\sim$ 168 degrees) it might be difficult to detect these parameters
in microscopic measurements of the speed of light performed in a
laboratory. The link between the two concepts depends on the adopted
model for the vacuum. The point of view adopted so far corresponds
to consider the vacuum as some kind of fluid in a state of regular,
laminar motion. In these conditions global and local properties of
the flow coincide.

We believe that, without fully understanding the nature of that
substratum that we call physical vacuum, one should instead keep a
more open mind. As discussed in Sect.2, the physical vacuum might be
similar to a form of turbulent ether, an idea which is deep rooted
in basic foundational aspects of both quantum theory and relativity
and finds additional motivations in those representations of the
vacuum as a form of `space-time foam' which indeed resembles a
turbulent fluid. In this case, global and local velocity fields
might be very different and there could be forms of random signals
that have a genuine physical origin. For instance, by combining the
point of view of ref.\cite{gerg}, where gravity is considered a
long-wavelength phenomenon which emerges from a space-time which is
fundamentally flat at very short distances, with the idea of a
turbulent ether, one arrives to an instantaneous stochastic signal
of typical magnitude $10^{-15}$ which could fit very well with the
present experimental data.

For this reason, after reviewing in Sects.3$-$4 the general
theoretical framework and the basics of the modern ether-drift
experiments, we have presented in Sect.5 a numerical simulation of
the possible effects that one might expect in a simple model where,
at small scales, vacuum turbulence appears statistically isotropic
and homogeneous. After subtracting the known forms of disturbances,
we have found that the observed distribution of the instantaneous
data requires a value $\tilde v\sim 332$ km/s of the scalar velocity
parameter which characterizes the fluctuations. Remarkably, this has
a definite counter part in the known Earth's cosmic motion with
respect to the CMB. In fact, it corresponds exactly to the average
projection of the Earth's velocity in the interferometer's plane for
an apparatus placed at the latitude of the laboratories in
Berlin-D\"usseldorf. However, by the very nature of the model, this
correspondence with the global Earth's motion is only valid at the
level of statistical distributions and is not detectable from the
naive time dependence of the data. We have also found that,
differently from trivial thermal noise, the stochastic signal of an
underlying turbulent vacuum should exhibit a transition from
non-gaussian to gaussian distributions of the data by increasing the
averaging time, in agreement with analogous phenomena observed in
turbulent flows. Furthermore, the typical instantaneous signal
should be about 100 times larger than the short-term stability, a
few $10^{-18}$, which is expected with the forthcoming generation of
cryogenic experiments \cite{upgrading}. A confirmation of these
predictions would represent compelling evidence for an
unconventional form of ether-drift with non-trivial implications for
our understanding of both gravity and relativity.

We emphasize that the existence of a genuine ether drift could have
other non-trivial consequences. In fact, in agreement with the
intuitive notion of an ether wind, it would mean that all physical
systems are exposed to a tiny energy flux, an effect that, in
principle, can induce forms of spontaneous self-organization in
matter \cite{prigogine}. In slightly different terms, the detection
of a stochastic drift implies that not all possible effects of the
underlying vacuum state get re-absorbed into the basic parameters of
the physical theory but there remains a weak, residual form of
`noise'. In principle, this fundamental noise, intrinsic to natural
phenomena (`objective noise' \cite{grigolini}), could be crucial. In
fact it has becoming more and more evident that, thanks to the
presence of noise, many classical and quantum systems can increase
their efficiency and evolve toward a more ordered behaviour compared
to the fictitious situation where spatial and/or temporal randomness
were absent \cite{gamma2} (see e.g. photosynthesis in sulphur
bacteria \cite{caruso}, protein crystallization \cite{frenkel},
noise enhanced stability \cite{spagnolo} or stochastic resonance
\cite{gamma1}).

In this sense, the outcome of ether-drift experiments could
determine a new framework where long-range correlations, complexity
and even life, might be thought as ultimately emerging, at higher
physical levels, from underlying dynamical processes. Specifically,
the idea of a turbulent ether introduces a peculiar element of
statistical physics, namely those `fat-tailed' Probability Density
Functions, characteristic of turbulent flows at short time scales
\cite{beck,sreenivasan}, that also characterize many complex systems
(see e.g. \cite{beck-cohen1,beck-cohen2,tsallis1,tsallis}) .

\vfill\eject

\end{document}